\documentclass[aps,pre,reprint,groupedaddress,A4]{revtex4-2}

\usepackage[utf8]{inputenc}
\usepackage[english,russian,main=english]{babel}

\usepackage{color}

\usepackage{graphicx}
\usepackage{amsmath}

\usepackage{tikz}
\usetikzlibrary{arrows}
\usepackage{pgfplots}

\usepackage{titlesec}

\renewcommand\vec{\mathbf}

\begin{document}


\title{Lagrangian for Navier-Stokes equations of motion: SDPD approach}


\author{Tetiana Kornylova}
\author{Anna Shokhina}
\affiliation{Department of Mathematics, Odessa I. I. Mechnikov National University, Odessa, Ukraine}
\author{Timothy Nerukh}
\affiliation{School of Mathematical Sciences, University of Southampton, Southampton SO17 1BJ, UK}
\author{Dmitry Nerukh}
\email{D.Nerukh@aston.ac.uk}
\affiliation{Department of Mathematics, Aston University, Birmingham, B4 7ET, UK}


\date{\today}

\begin{abstract}
The conditions necessary and sufficient for the Smoothed Dissipative Particle 
Dynamics (SDPD) equations of motion to have a Lagrangian that can be used for 
deriving these equations of motion, the Helmholtz conditions, are obtained and 
analysed. They show that for a finite number of SDPD particles the conditions 
are not satisfied; hence, the SDPD equations of motion cannot be obtained 
using the classical Euler-Lagrange equation approach.  However, when the macroscopic 
limit is considered, that is when the number of particles tends to infinity, 
the conditions are satisfied, thus providing the conceptual possibility of 
obtaining the Navier-Stokes equations from the principle of least action. 
\end{abstract}


\maketitle

\section{Introduction}

Obtaining hydrodynamic equations of motion from the fundamental Action 
Principle is a largely unexplored area of research.  Even though the 
generalisation of classical Action Principle for particles to continuous fields 
is known for some time \cite{Goldstein}, attempts to take into account energy 
dissipation are rare and only recently such Lagrangians for fluid dynamics equations 
of motion are reported \cite{Bennett2006,Materassi2015,Musicki2016}. An 
advantage provided by such Lagrangian approach is in natural connection between 
the dynamics of discrete particles and continuous fields best suitable for 
describing fluids at macroscale. When particles represent atoms (even at the 
classical approximation) this connection is the physical foundation of the 
interaction between the scales representing multiscale, multiphysics description of 
liquids, which is an active area of research.

To the best of our knowledge Lagrangians that lead to classical Navier-Stokes
(NS) equations of hydrodynamics are unknown. Here we take the first step in the
direction of finding such Lagrangians, namely we seek to answer the question if
such Lagrangians exist. Our  approach consists of first considering the discrete
approximation of NS equations, the Smoothed Dissipative Particles Dynamics
(SDPD) model \cite{Espanol2003}, that converges to continuum NS equations in the
limit of infinite number of particles.  There are mathematical conditions,
attributed to Helmholtz, that if fulfilled guarantee the existence of the
Lagrangian.  We check these conditions for SDPD equations of motion and then
analyse their behaviour for the macroscopic limit.

Our results show that the conditions are not satisfied for SDPD equations with
finite number of particles.  However, we demonstrate that the discrepancy decreases with increasing
the number of particles.  Also, in the limit of
infinite number of particles the conditions are satisfied, thus providing an 
approach for obtaining a Lagrangian for the Navier-Stokes equations.

\section{Theory}

Mathematically, finding a Lagrangian for a system of equations of motion amounts 
to solving the inverse problem of the calculus of  
variations \cite{Vujanovic1989}. 
It is known that for a system of $n$ given differential equations 
\begin{equation}\label{eq:H}
H_i(t,q_i,\dot{q}_i,\ddot{q}_i)=0 ,
\end{equation}
the necessary and sufficient condition that the 
system is derivable from a Lagrangian is that the equations of variation of 
$H_i$ form a self-adjoint system.  The conditions under which the system of 
variations is self-adjoint are attributed to Helmholtz \cite{Vujanovic1989}:
\begin{equation}\label{HelmholtzCondition1}
\frac{\partial H_i}{\partial \ddot{q}_j} = \frac{\partial H_j}{\partial \ddot{q}_i}, 
\end{equation}
\begin{equation}\label{HelmholtzCondition2}
\frac{\partial H_j}{\partial \dot{q}_i} + \frac{\partial H_i}{\partial \dot{q}_j} = 
2 \frac{d}{dt} \left( \frac{\partial H_i}{\partial \ddot{q}_j}  
\right), 
\end{equation}
\begin{equation}\label{HelmholtzCondition3}
\frac{\partial H_j}{\partial q_i}  = \frac{\partial H_i}{\partial q_j} - 
\frac{d}{dt} \left( \frac{\partial H_i}{\partial \dot{q}_j}  \right) + 
\frac{d^2}{dt^2} \left( \frac{\partial H_i}{\partial \ddot{q}_j}  \right),
\end{equation}
$ \forall i, j = 1, \dots ,n.$
If these conditions are satisfied, $H_i$ must take the form 
$H_i=M_i+P_{ij}\ddot{q}_j$, where $M_i$ and $P_{ij}$ are functions related to 
each other (see \cite{Vujanovic1989})  
and the Lagrangian can be constructed from the functions $M_i$ and $P_{ij}$.  
Importantly, a Lagrangian constructed this way does not necessarily have the 
usual physical meaning of the difference between the kinetic and potential 
energies.  Rather, it is an abstract mathematical function that, when used in 
Euler-Lagrange equation, produces the required equations of motion.

In the SDPD framework the fluid is represented by a set of $N$ particles, each of
which is considered as a macroscopic thermodynamic system of constant mass $m_i$
\cite{Espanol2003}.  Particles are described by their positions $\vec{r}_i$,
velocities $\vec{v}_i$, and entropy $S_i$ (or internal energy).  The particle's volume $\nu_i$ is
defined as its inverse number density $d_i$:
 
\begin{equation}\label{density}
     \frac{1}{\nu_i} = d_i = \sum \limits_j W(|\vec{r}_i - \vec{r}_j|),
\end{equation}
where $W (r, h) = W (| r |, h)$ is a pairwise bell-shaped interpolation 
function of compact support $h$ (the kernel). Various forms of $W$ exist, we used the Lucy 
function 
\begin{equation}\label{LucyFunc}
     W(r) = \frac{105}{16 \pi h^3} \left( 1 + 3 \frac{r}{h}\right) \left( 1 - 
\frac{r}{h}\right)^3.
\end{equation}
The gradient of $W$ defines the function $F (r)$ as $\nabla W(r) = -r F(r), \:  F(r) \ge 0$,
which for the Lucy kernel has the form
\begin{equation}\label{Ffunc}
     F(r) = \frac{315}{4 \pi h^5} \left( 1 - \frac{r}{h} \right)^2.
\end{equation}

Using the auxiliary quantities $\vec{r}_{ij} = \vec{r}_i - \vec{r}_j$,
$\vec{v}_{ij} = \vec{v}_i - \vec{v}_j$, and $F_{ij} = F(|\vec{r}_{ij}|)$ the
SDPD equations of motion read
\begin{eqnarray}
\dot{\vec{r}}_i & = & \vec{v}_i, 
\nonumber\\
m \dot{\vec{v}}_i & = &\sum \limits_j \left[ \frac{P_i}{d_i^2} + 
\frac{P_j}{d_j^2}\right] F_{ij} \vec{r}_{ij} - \left( \frac{5 \eta}{3} - \xi \right) 
\sum \limits_j \frac{F_{ij}}{d_i d_j} \vec{v}_{ij} - 
\nonumber\\
& & 5 \left( \xi + \frac{\eta}{3} 
\right) \sum \limits_j \frac{F_{ij}}{d_i d_j} \vec{e}_{ij} \vec{e}_{ij} \cdot \vec{v}_{ij}, 
\nonumber\\
T_i \dot{S_i} & = & \phi_i - 2 \kappa \sum \limits_j \frac{F_{ij}}{d_i d_j} 
(T_i - T_j), \label{eq:SDPD}
\end{eqnarray}
where
\begin{eqnarray}
     \vec{e}_{ij} & = & \frac{\vec{r}_i - \vec{r}_j}{|\vec{r}_i - \vec{r}_j|},
\nonumber\\
    \phi_i & = & \left( \frac{5 \eta}{6} - \frac{\xi}{2} \right) \sum \limits_j \frac{F_{ij}}{d_i d_j} \vec{v}_{ij}^2 + 
\nonumber\\
& & \frac{5}{2} \left( \xi + \frac{\eta}{3} \right) \sum \limits_j 
\frac{F_{ij}}{d_i d_j} \left( \vec{e}_{ij} \cdot \vec{v}_{ij}\right)^2,
\nonumber
\end{eqnarray}
$T_i$ and $P_i$ are particles' temperature and pressure (calculated using 
phenomenological equations of state), $\eta$ and $\xi$ are viscosities, 
$\kappa$ is the thermal conductivity \cite{Espanol2003}.

\vspace{-1mm}

We have used the following approximations in our SDPD model: (i) temperature 
was constant and equal for all particles, (ii) the shear viscosity $\eta$ was 
constant and equal for all particles, (iii) the bulk viscosity $\xi$ was negligible, 
(iv) pressure was calculated using the Tait equation of state
\begin{equation*}
P_i = B \left[ \left(\cfrac{\rho_i}{\rho_0}\right)^{\gamma} - 1 \right],
\end{equation*}
where $\gamma = 7$, $B = \cfrac{c^2_0 \rho_0}{\gamma},$ $\rho_0$ is the control 
density, $\rho_i=m_i d_i$ is the physical mass density, and $c_0$ is the control speed of sound.

The equation for the entropy in (\ref{eq:SDPD}) is often considered decoupled
from the equations for the position and the velocity to a good approximation,
depending on the equations of state for $P$ and $T$. Therefore, we do not take
it into account here as our model equations of state do not depend on $S$. After
substituting $\dot{\vec{r}}_i$ for $\vec{v}_i$ in the second equation in
(\ref{eq:SDPD}) the system (\ref{eq:H}) for independent variables $\vec{r}_i$,
$x$, $y$, and $z$ components of which represent variables $q_i$, become (note,
that indexes $i$ refer to different variables here: the particle number for
$\vec{r}_i$ and the degree of freedom for $q_i$)
\begin{widetext}
\begin{eqnarray}\label{eq:H_SDPD}
   H_{i} &=& C \sum_{j} \left[ \left(  B \left( \frac{m_i}{\rho_{0}} 
\right)^{7} d_{i}^{5}- Bd_{i}^{-2} +  B \left( \frac{m_j}{\rho_{0}} 
\right)^{7} d_{j}^{5}- Bd_{j}^{-2}\right) K_{ij}\vec{r}_{ij}\right] - \nonumber\\
&&C \sum_{j} \left[ K_{ij} d_{i}^{-1} d_{j}^{-1} \left( A \cdot \dot{\vec{r}}_{ij} + D \vec{r}_{ij} 
\left( \vec{r}_{ij} \cdot \dot{\vec{r}}_{ij} \right) \left( \left| \vec{r} \right|_{ij} \right)^{-2} 
\right)\right] 
 - m_i \ddot{\vec{r}}_{i},
\end{eqnarray}
\end{widetext}
where $A = \left( \frac{5 \eta}{3} - \xi \right)$, 
$D =  \left( \xi + 
\frac{\eta}{3} \right)$, $C =  \frac{315}{4 \pi h^{5}}$ are constants and $K_{ij}= \left(1- 
\frac{\left | \vec{r} \right |_{ij}}{h} \right)^{2}$, $d_{i} = \sum_{j}  \frac{105}{16 \pi h^{3}} \left ( 
1+3\frac{\left | \vec{r} \right |_{ij}}{h} \right )\left(1- \frac{\left | \vec{r} \right 
|_{ij}}{h} \right)^{3}$. 

\vspace{-3mm}

Therefore, checking the Helmholtz conditions for SDPD equations amounts to
checking equations (\ref{HelmholtzCondition1})-(\ref{HelmholtzCondition3}) for
$H_i$ functions defined by (\ref{eq:H_SDPD}).

\vspace{-3mm}

\subsection{First Helmholtz condition}

\vspace{-3mm}

The first condition (\ref{HelmholtzCondition1}) is satisfied for all variables 
as the left and the right hand sides of the equation are 0 for $i\ne j$ since the equation 
of motion for particle $i$ does not depend on the second time derivative of the 
coordinate of a different particle $j$, while the case $i=j$ is trivially 
satisfied.

As we show below, the second (\ref{HelmholtzCondition2}) and the third 
(\ref{HelmholtzCondition3}) conditions are not satisfied and we  
analyse the residues as we are interested in the behaviour of these residues 
when the number of particles tends to infinity.

\subsection{Second Helmholtz condition}
\label{sec:H2}

In our case, the second Helmholtz condition becomes:

\begin{equation}\label{cond2}
     \frac{\partial H_j}{\partial \dot{q_i}} + \frac{\partial H_i}{\partial \dot{q_j}} = 0,
\end{equation}
for $i, j = 1, \dots ,n=3N.$ There are four possibilities for this condition 
after assigning the particles' coordinates to variables $q_i$ and corresponding 
functions $H_i$ (again, indexes $i$ have different meaning depending if they 
refer to the degree of freedom as in (\ref{HelmholtzCondition2}) and 
(\ref{cond2}) or to the particle number as in (\ref{eq:SDPD})): 
 \begin{enumerate} 
 \item Particle $i$ with respect to itself in the same coordinate $x$ 
($y$, $z$):
 \begin{equation}\label{eq:HC2_1}
      \frac{\partial H_{xi}}{\partial \dot{x_i}} + \frac{\partial 
H_{xi}}{\partial \dot{x_i}} = \sum_{j}T_{ij}d_{i}^{-1}d_{j}^{-1}Q^{x}_{ij}.
 \end{equation}
 \item Particle $i$ with respect to itself in different coordinates $x$ and $y$ 
(and other pairs of coordinates):
 \begin{equation}\label{eq:HC2_2}
      \frac{\partial H_{xi}}{\partial \dot{y_i}} + \frac{\partial 
H_{yi}}{\partial \dot{x_i}} = \sum_{j}T_{ij}d_{i}^{-1}d_{j}^{-1}Q^{xy}_{ij}.
 \end{equation}
 \item Particle  $i$ with respect to a different particle $j$ in the same coordinate $x$
($y$, $z$):
 \begin{equation}\label{eq:HC2_3}
      \frac{\partial H_{xi}}{\partial \dot{x_j}} + \frac{\partial 
H_{xj}}{\partial \dot{x_i}} = T_{ij}d_{i}^{-1}d_{j}^{-1}Q^{x}_{ij}.
 \end{equation}
 
 \item Particle  $i$ with respect to a different particle $j$ in different 
coordinates $x$ and $y$:
\begin{equation}\label{eq:HC2_4}
      \frac{\partial H_{xi}}{\partial \dot{y_j}} + \frac{\partial 
H_{yj}}{\partial \dot{x_i}} = \frac{\partial H_{xj}}{\partial \dot{y_i}} + 
\frac{\partial H_{yi}}{\partial \dot{x_j}} =  
T_{ij}d_{i}^{-1}d_{j}^{-1}Q^{xy}_{ij}.
 \end{equation}
 \end{enumerate}
 
The following variables were used:

\begin{align}
    &  T_{ij}= CK_{ij}m_{i}m_{j}, \\
    &  Q^{x}_{ij}= A + D x_{ij} \left( \left| r \right|_{ij} \right)^{-2}, \\
    &   Q^{xy}_{ij}=  D x_{ij} y_{ij} \left( \left| r \right|_{ij} \right)^{-2}. 
\end{align}
 
\subsection{Third Helmholtz condition}
\label{sec:H3}

Our functions $H_i$ lead to the following third Helmholtz condition:

\begin{equation}\label{cond3}
     \frac{\partial H_j}{\partial q_i} - \frac{\partial H_i}{\partial q_j} + \frac{\partial }{\partial t}\left(\frac{\partial H_i}{\partial \dot{q_j}}\right) = 0,
\end{equation}
for $i, j = 1, \dots ,n=3N$ with four realisations:
\begin{enumerate} 
\item Particle $i$ with respect to itself in the same coordinate $x$ 
($y$, $z$):
\begin{equation}\label{eq:HC3_1}
\frac{\partial H_{xi}}{\partial {x_i}} -  \frac{\partial H_{xi}}{\partial {x_i}} +  \frac{\partial }{\partial t}\left(  \frac{\partial H_{xi}}{\partial \dot{x_i}}\right) = R^{x}_{ij}, 
\end{equation}
variable $R^{x}_{ij}$ is defined in appendix \ref{App}.

\begin{widetext}
  
\item  Particle $i$ with respect to itself in different coordinates $x$ and $y$ 
(and other pairs of coordinates):
\begin{multline}\label{eq:HC3_2_1}
 \frac{\partial H_{xi}}{\partial {y_i}} -  \frac{\partial H_{yi}}{\partial {x_i}} +  \frac{\partial }{\partial t}\left(  \frac{\partial H_{yi}}{\partial \dot{x_i}}\right) = \\ = 
 \sum_{j} \left[ T_{ij} \left (\left(L_{i} y_{ij} + d_{i}^{-2}d_{j}^{-1}S_{ij}^{y} \right)\left(\sum_{k}CK_{ik}x_{ik}\right) - \right.\right. \\
  \left.\left. -\left(L_{i} x_{ij} + d_{i}^{-2}d_{j}^{-1}S_{ij}^{x} \right) \left(\sum_{k}CK_{ik}y_{ik}\right)  \right) + \right.  \\
+ \left.T_{ij}d_{i}^{-1}d_{j}^{-1}\left[v_{ij}x_{ij}y_{ij} - v_{ij}y_{ij}x_{ij} \right]\right. \\ 
\left.\left[  \frac{2}{ h\left|r\right|_{ij} \left(1 - 
\frac{\left|r\right|_{ij}}{h}\right)} A - d_{j}^{-1}K_{ij}A  + 
D\left(\left|r\right|_{ij}\right)^{-2}\right]\right] + R^{xy}_{ij}.
 \end{multline}
The expression for $\frac{\partial H_{yi}}{\partial {x_i}} -  \frac{\partial 
H_{xi}}{\partial {y_i}} +  \frac{\partial }{\partial t}\left(  \frac{\partial 
H_{xi}}{\partial \dot{y_i}}\right)$ is almost the same with two signs reverted, 
see appendix \ref{App}, equation (\ref{eq:HC3_2_2}).

\item Particle $i$ with respect to a different particle $p$ in the same coordinate $x$
($y$, $z$):
\begin{multline}\label{eq:HC3_3_1}
 \frac{\partial H_{xi}}{\partial {x_p}} -  \frac{\partial H_{xp}}{\partial {x_i}} +  \frac{\partial }{\partial t}\left(  \frac{\partial H_{xp}}{\partial \dot{x_i}}\right) =  \\ = 
 C \sum_{j} \left[ T_{ij} \left( K_{ip}x_{ip} \left[  L_{i}x_{ij}  +     d_{i}^{-2}d_{j}^{-1}S^{x}_{ij} \right] -  K_{pj}x_{pj}\left[    L_{j}x_{ij}+ d_{i}^{-1}d_{j}^{-2}S^{x}_{ij} \right]   \right) + \right. \\
\left. + T_{ip} K_{ij}x_{ij} \left( L_{i}x_{ip} + d_{i}^{-2}d_{p}^{-1}S^{x}_{ip} \right) \right]  + \\
 + C \sum_{l} \left(  T_{pl} \left[  K_{ip}x_{ip} \left(  L_{p}x_{pl} +  d_{p}^{-2}d_{l}^{-1}S^{x}_{pl}      \right) + K_{il}x_{il} \left(   L_{l}x_{pl} +   d_{p}^{-1}d_{l}^{-2}S^{x}_{pl}   \right)   \right] + \right.\\
 + \left.T_{ip} K_{pl}x_{pl}  \left( L_{p}x_{ip} + d_{i}^{-1}d_{p}^{-2}S^{x}_{ip} 
 \right)\right) +  R^{x}_{ip} .
\end{multline}
The expression for $\frac{\partial H_{xp}}{\partial {x_i}} -  \frac{\partial 
H_{xi}}{\partial {x_p}} +  \frac{\partial }{\partial t}\left(  \frac{\partial 
H_{xi}}{\partial \dot{x_p}}\right)$ is almost the same with two signs at the summations reversed, 
see appendix \ref{App}, equation (\ref{eq:HC3_3_2}).

\item Particle  $i$ with respect to a different particle $p$ in different 
coordinates $x$ and $y$:
\begin{multline}\label{eq:HC3_4_1}
\frac{\partial H_{xi}}{\partial {y_p}} -  \frac{\partial H_{yp}}{\partial {x_i}} +  \frac{\partial }{\partial t}\left(  \frac{\partial H_{yp}}{\partial \dot{x_i}}\right) =\\
= T_{ip} d_{i}^{-1}d_{p}^{-1}\left[ v_{ip}y_{ip}x_{ip}-v_{ip}x_{ip}y_{ip}\right]\left[\frac{2}{h\left|r\right|_{ip} \left(1 - \frac{\left|r\right|_{ip}}{h}\right)} A + D\left(\left|r\right|_{ip}\right)^{-2}\right] + \\
+ C\sum_{j} \left[ T_{ij} \left(K_{ip}y_{ip}\left[ L_{i}x_{ij} + d_{i}^{-2}d_{j}^{-1}S_{ij}^{x}\right] - K_{pj}y_{pj}\left[ L_{j}x_{ij} + d_{i}^{-1}d_{j}^{-2}S_{ij}^{x}\right] \right) - \right. \\
 \left. - T_{ip} K_{ij}x_{ij} \left(L_{i} y_{ip} + d_{i}^{-2}d_{p}^{-1}S_{ip}^{y} \right) \right] + \\
 + C\sum_{l} \left[ T_{pl} \left(K_{ip}x_{ip}\left[ L_{p}y_{pl} + d_{p}^{-2}d_{l}^{-1}S_{pl}^{y}\right]  + K_{il}x_{il}\left[ L_{l}y_{pl} + d_{p}^{-1}d_{l}^{-2}S_{pl}^{y}\right]\right) - \right. \\
 \left. - T_{ip} K_{pl}y_{pl}\left(L_{p} x_{ip} + 
d_{i}^{-1}d_{p}^{-2}S_{ip}^{x} \right) \right] +  R^{xy}_{ip}.
 \end{multline} 
The analogous expressions for the other combinations of the coordinates as well
as definitions of the auxiliary variables are given in appendix \ref{App}, equations
(\ref{eq:HC3_4_2}-\ref{eq:last}).

\end{widetext}
 
\end{enumerate}

As the right hand sides of equations (\ref{eq:HC2_1}-\ref{eq:HC2_4}), 
(\ref{eq:HC3_1}-\ref{eq:HC3_4_1}) 
are clearly not 0, the second and the third Helmholtz conditions are not 
satisfied for a system of finite number of SDPD particles.

\section{The macroscopic limit leading to continuous Navier-Stokes system}

Even though the second and the third conditions are not satisfied, it is 
interesting to investigate how large the discrepancy is.  In other words, what 
is the value of the right hand sides of the conditions, which should be zero 
if they were satisfied.  This is an important question as the SDPD equations 
are a discrete approximation of the continuous Navier-Stokes equations.  It is known
that the SDPD equations converge to Navier-Stokes equation in the limit of infinite number 
of particles (see below for details).

The original SDPD equations have two parts: deterministic and stochastic 
\cite{Espanol2009}. The deterministic part, eq. (3) in  \cite{Espanol2009}, 
defines the macroscopic dynamics and does not depend on the spatial and 
temporal scales (``scale-free''), that is, it is independent of the volume of 
particles $\nu_i$. On the contrary, the stochastic part,  eq. (4) in  
\cite{Espanol2009}, defines the scale and it is inversely proportional to the 
particles' volume.  In the large particles limit, the stochastic part is 
negligible.  Therefore, in the macroscopic limit, only the deterministic part 
of SDPD equations needs to be considered, which are our equations 
(\ref{eq:SDPD}). 

\subsection{How particles converge to hydrodynamic fields}

On the other hand, the SDPD equations are the discrete approximation for 
continuous Navier-Stokes equations describing the evolution of hydrodynamic 
fields $\rho$, $\vec{v}$, and $S$.  As explained in \cite{Lucy1977}, in order 
to obtain the continuous limit of the discrete approximation, the number of 
particles $N$ should tend to infinity and, at the same time, the width of the 
kernel's support $h$ should tend to $0$.  In this limit the set of $N$ SDPD 
equations tends to three Navier-Stokes equations describing the continuous 
hydrodynamic fields.   

We, therefore, need to analyse (or define) how the quantities on the right hand 
side of equations (\ref{cond2}-\ref{eq:HC3_4_1}) depend on $N$ and take the 
limit $N\to \infty$, making sure that in this limit also $h\to 0$. 

\subsection{The Lucy kernel function and particle's number density}

The most involved quantity to estimate is the volume of the particles as it is defined through the summation involving the kernel function (\ref{density}).  To analyse its dependence on $N$ let us first consider the Lucy function written in the form
\begin{equation}\label{LucyW1}
     W(r;h) = \frac{105}{16 \pi} W_1(r;h),
\end{equation}
where
\begin{equation}\label{eq:W1}
    W_1(r;h)=\frac{1}{h^3}\left( 1 + 3 \frac{r}{h}\right) \left( 1 - \frac{r}{h}\right)^3
\end{equation}
and $W_1(r)=0$ for $r<0$ or $r>h$, Fig. \ref{fig:W1}.

\begin{figure}
\begin{tikzpicture}[
    axis/.style={->, >=stealth'},
    line/.style={thick,rounded corners=8pt},
    ]
\begin{axis}[
    width=.43\textwidth,
    height=.2\textheight,
    xmin=0, xmax=1.7,
    ymin=0, ymax=0.35,
    xtick={0, 1.5},
    xticklabels={$0$, $h$},
    ytick={0, 0.296},
    yticklabels={$0$, $\frac{1}{h^3}$},
    axis lines = left,
    xlabel = \(r\),
    ylabel = {\(W_1(r)\)},
]
    \addplot[domain=0:2,samples=50]{1/1.5^3*(1+3*x/1.5)*(1-x/1.5)^3};
\end{axis}
\end{tikzpicture}
\caption{\label{fig:W1}
The form of the function $W_1$}
\end{figure}

From (\ref{density}), the number density $d_i$ of a particle is equal to
\begin{equation}\label{eq:d_W1}
    \frac{1}{\nu_i}=d_i=\frac{105}{16 \pi}\sum_{j=1}^{K} W_1(|\vec{r}_{ij}|),
\end{equation}
where $K$ is the number of particles inside the sphere of radius $h$.  $K$ is approximately equal to the fraction of the volume of this sphere in the total volume of the system $V_0$:
\begin{equation}\label{eq:K}
    K=\frac{\nu_i}{V_0}N=\frac{4\pi}{3V_0}h^3N.
\end{equation}
In order to estimate the value of the sum in (\ref{eq:d_W1}) 
we need to estimate how the distances $|\vec{r}_{ij}|$ are distributed.

\begin{figure}
\includegraphics[width=\linewidth]{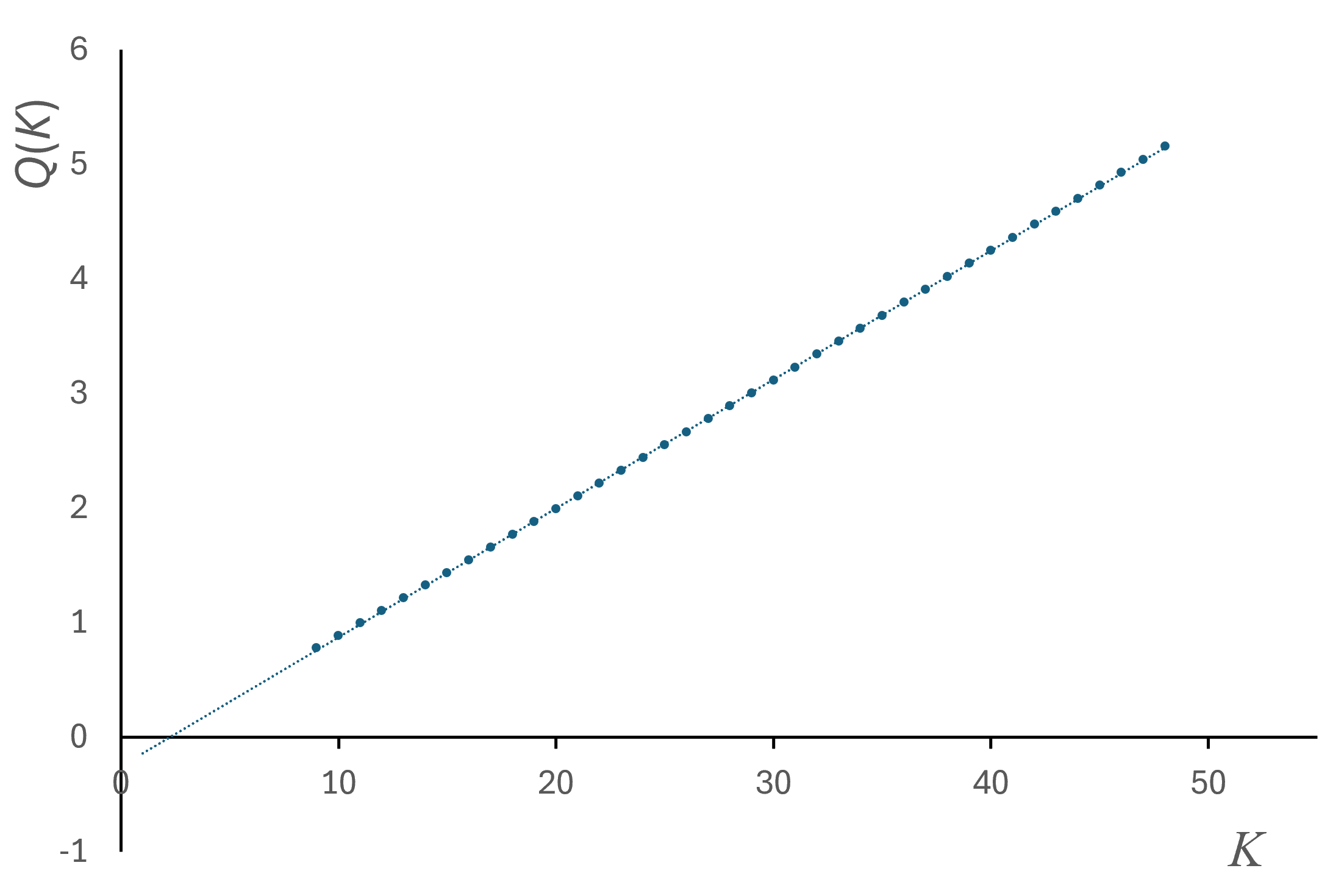}
\caption{\label{fig:Q(K)}
The linear dependence of function $Q$ on $K$
}
\end{figure}

In the assumption that the particles are distributed uniformly in 3D space, the 
\emph{distances} $r$ from the centre of a sphere of radius $R$ are distributed 
according to the cumulative distribution function (CDF) 
$F(r)=\left(\frac{r}{R}\right)^3$. The distribution of distances $r$ are, then, 
obtained using inverse transform sampling, where a uniformly distributed values 
$U$ are transformed by the inverse CDF: $r=F^{-1}(U)$, which in our case is 
equal to $r=RF^{\frac{1}{3}}$. Taking an equidistant set of points 
$\frac{j}{K}$ as a set of uniformly distributed points, the value of the distances becomes 
\begin{equation*}
    |\vec{r}_{ij}|=\left(\frac{j}{K}\right)^{\frac{1}{3}}h.
\end{equation*}
Substituting this value into the sum of (\ref{eq:d_W1}) we obtain the following expression
\begin{equation*}
    \frac{1}{h^3}\sum_{j=1}^K \left[1-3\left(\frac{j}{K}\right)^{\frac{1}{3}}\right]\left[1-\left(\frac{j}{K}\right)^{\frac{1}{3}}\right]^{3} = \frac{1}{h^3}Q(K).
\end{equation*}
Function $Q$ is very close to a linear function for all $K>8$, Fig. \ref{fig:Q(K)}, and can be approximated as $Q(K)\approx 0.11K$.

Substituting this and the value for $K$ from (\ref{eq:K}), we obtain the estimation for the number density $d_i$ as a function of $N$:
\begin{equation}
    d_i=\frac{105}{16\pi}\frac{1}{h^3}Q(K)\approx \frac{105\cdot 0.11}{12V_0}N=EN
\end{equation}
with $E$ constant.

\subsection{The macroscopic limit}

In order to estimate the macroscopic limit, what remains is to estimate the distance $x_{ij}$, which is simply the $x$-component of the average distance between the particles and, as such, can be assumed to be equal to $\left(\frac{V_0}{N}\right)^{\frac{1}{3}}$ multiplied by $j$ with plus or minus sign depending on the direction of the vector $\vec{r}_{ij}=\vec{r}_i-\vec{r}_j$:
\begin{equation*}
    x_{ij}=\pm\left(\frac{V_0}{N}\right)^{\frac{1}{3}}j.
\end{equation*}
The mass of particles is equal to $m_i=\frac{M_0}{N}$, where $M_0$ is the total mass of the system.

Finally, we need to define how the kernel support $h$ tends to zero with $N\to 
\infty$.  A reasonable assumption here is to require the radius of the sphere 
$h$ to be such, that there is always a fixed number of particles $K_0$ 
inside the sphere. Hence, from (\ref{eq:K}), the value of $h$ becomes 
\begin{equation}
    h=\left(\frac{K_0}{N}\frac{3V_0}{4\pi}\right)^\frac{1}{3}=GN^{-\frac{1}{3}},
\end{equation}
with $G$ constant.

Combining these results, the right hand side of equation (\ref{eq:HC2_1}) becomes
\begin{align}
 & \frac{\partial H_{xi}}{\partial \dot{x_i}}  + \frac{\partial 
H_{xi}}{\partial \dot{x_i}} =   \\
 & \frac{315}{4\pi}\sum_{j}^{K_0}\frac{1}{h^5}\left(1-\frac{|\vec{r}|_{ij}}{h}\right)^2 m_i m_j
\frac{1}{d_i}\frac{1}{d_j}\left(A+Dx_{ij}\frac{1}{|\vec{r}|_{ij}^2}\right) = \nonumber \\
 & \frac{315}{4\pi}\sum_{j}^{K_0} \left(GN^{-\frac{1}{3}}\right)^{-5}
 \left(1-\left(\frac{j}{K_0}\right)^{\frac{1}{3}}\right)^2 M_0^2(NE)^{-2}(N)^{-2} \nonumber \\
 & \left(A\pm DV_0^{\frac{1}{3}}N^{-\frac{1}{3}}j\left(\frac{j}{K_0}\right)^{-\frac{2}{3}}\left(GN^{-\frac{1}{3}}\right)^{-2}\right). \nonumber
\end{align}
In this expression, all terms with the exception of $N$ do not grow larger than $K_0$.  
Ignoring constants, the dependence on $N$ of the terms under the sum is  
\begin{equation}
    \frac{\partial H_{xi}}{\partial \dot{x_i}}  + \frac{\partial 
H_{xi}}{\partial \dot{x_i}} \sim \sum_{j}^{K_0} \left(N^{-\frac{7}{3}}\pm N^{-2}\right),
\end{equation}
which tends to $0$ in the limit $N\to \infty$.  

Similar analysis of the other three possibilities for the second Helmholtz 
condition, section \ref{sec:H2}, leads to analogous expressions that depend on 
negative powers of $N$: $N^{-\frac{7}{3}}$ and $N^{-\frac{7}{3}}\pm N^{-2}$ (Appendix \ref{app:limits}), thus tending to $0$ in the macroscopic limit.

For the third condition, section \ref{sec:H3}, we also needed to make an 
additional assumption on how the velocity difference $\vec{v}_{ij}$ tends to 
$0$ with the increase of the number of particles $N$, and, hence, the decrease 
of the distance between the particles. Assuming the same behaviour as for 
$\vec{r}_{ij}$, that is $\vec{v}_{ij}\sim N^{-\frac{1}{3}}$, similar results 
for the right hand sides of equations (\ref{eq:HC3_2_1}-\ref{eq:HC3_4_1}) are 
obtained, that is the dependence on negative powers of $N$, leading to their 
vanishing in the $N\to \infty$ limit (Appendix \ref{app:limits}).

\section{Simulation details and numerical results}

To confirm the analysis above, we have also performed numerical estimation of 
the right hand sides of the conditions by numerically simulating an SDPD system with 
varying number of particles and kernel width $h$ (keeping all other parameters the same).  Our results 
show that with growing $N$ and decreasing $h$, the right hand sides (the deviation from 0) become 
smaller, thus confirming the behaviour in the macroscopic limit.

We used an implementation for SDPD developed as a package for the popular 
Molecular Dynamics simulator LAMMPS \cite{Plimpton1995}.  The package 
{\tt USER-SDPD} is described in \cite{Jalalvand2020}. It simulates water-like 
liquid with approximations described by (\ref{eq:SDPD}) at mesoscopic 
scales where thermal fluctuations are small. The parameters of the simulated 
system are listed in table \ref{table}.  Cubic simulation box with periodic 
boundary conditions was used.

\begin{table}
\caption{Parameters of the simulated system
\label{table}}
\begin{ruledtabular}
\begin{tabular}{cll}
$\eta$ & shear viscosity & 1 $\text{pg}/(\mu \text{m} \cdot \mu \text{s})$  \\
$c_0$ & control sound velocity & 10 $\mu \text{m} / \mu \text{s}$ \\
$d_0$ & control density & 1 $\text{pg} / \mu \text{m}^3$  \\
$h$   & kernel width & 0.18 $\mu \text{m}$ \\
$T$   & temperature & 300 K   \\
$m_i$ & particle mass & 0.001 pg   \\
$dt$  & time step & $5\cdot 10^{-4}$ $\mu \text{s}$   \\
\end{tabular}
\end{ruledtabular}
\end{table}

The kernel support $h$ controls
how many neighbour particles were taken into account when calculating the
particle summations in SDPD formulas.  Thus, it should have a minimal value to ensure the stability of simulations.  Therefore, we
investigated how the values of calculated quantities changed with increasing
$h$.  We have found that in all cases the value of $h=0.18 \mu m$ was
sufficiently large for the calculated values to converge.

\begin{figure}[hb]
\includegraphics[width=\linewidth]{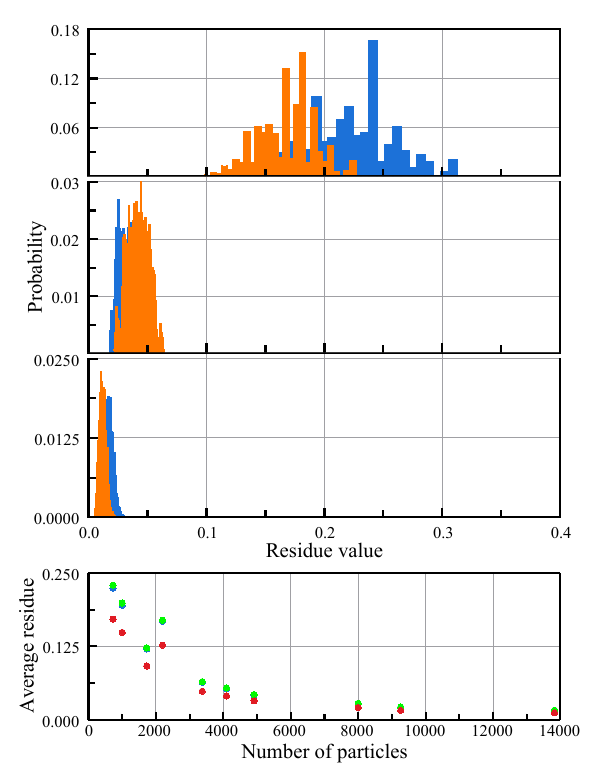}
\caption{Second Helmholtz condition residue as a function of the number of 
particles in the system -- right hand side of equations (\ref{eq:HC2_1}-\ref{eq:HC2_4}); top three plots show the distribution of the residue for all particles in the system, with the total number of particles 729, 4913, and 13824 for the top, middle and bottom plots, respectively (the blue histograms show residue (\ref{eq:HC2_1}), and the orange histograms show residue (\ref{eq:HC2_2}); their $x$ (top), $y$ (middle), and $z$ (bottom), components); the bottom plot shows the averaged over all particles in the system values of $x$ component of residue (\ref{eq:HC2_1}), blue, $y$ component of residue (\ref{eq:HC2_3}), green, and $z$ component of residue (\ref{eq:HC2_4}), red.
\label{fig:histograms}}
\end{figure}

The second condition in the form of equations (\ref{eq:HC2_1}-\ref{eq:HC2_4}) 
demonstrates clear convergence of the residue with increasing number of 
particles in the system, Fig.~\ref{fig:histograms}.
Several representative variables (different variants of the second condition and different components of the particles' coordinate) are shown as an illustration. The residues demonstrate clear tendency to lower values with increasing number of particles in the system.

We found that for the third condition in the form of equations 
(\ref{eq:HC3_1}-\ref{eq:HC3_4_1},\ref{eq:HC3_2_2}-\ref{eq:HC3_4_4}) the residue values tend towards zero in a similar manner.

\section{Conclusions}

In this paper we investigated if it is possible to derive Navier-Stokes 
hydrodynamic equations from a Lagrangian using the Euler-Lagrange equation.  
For this we 
analysed the three Helmholtz conditions necessary for a 
dynamical system to be derivable from a Lagrangian function.  As the dynamical 
system we used SDPD equations of motion that are discrete approximations of the 
Navier-Stokes equations that converge to them in the infinite number of 
particles and zero kernel function support limit.  We have found that the second and the third conditions are 
not satisfied for a finite number of particles, however the residues tend to 
zero with increasing number of particles. Also these residues tend to 0 in the analytical limit $N\to \infty$ and $h\to 0$. Thus, the continuous Navier-Stokes equations can be derived from a Lagrangian, at least in principle.

We are currently working on obtaining the explicit form of the Lagrangian in the macroscopic limit when the  
Helmholtz conditions are satisfied.

Finally, the presented results will be used as the basis for constructing 
hybrid particles, possessing simultaneously the properties of atoms and 
mesoscopic hydrodynamic particles, thus opening up the possibility of smooth 
transformation between physically distinct scales.

\titlespacing\section{0pt}{12pt plus 4pt minus 0pt}{0pt plus 2pt minus 2pt}
\titlespacing\subsection{0pt}{12pt plus 4pt minus 0pt}{0pt plus 2pt minus 2pt}
\titlespacing\subsubsection{0pt}{12pt plus 4pt minus 0pt}{0pt plus 2pt minus 2pt}

\begin{acknowledgments}
The authors acknowledge funding from Erasmus mobility programme ``2018 KA107 
International Credit Mobility'' for TK and AS visits to Aston University as well 
as H2020-MSCA-RISE-2018 programme, project AMR-TB, grant ID: 823922.
\end{acknowledgments}

\bibliography{refs}

\begin{appendix}

\begin{widetext}

\section{Expressions for the third Helmholtz condition}
\label{App}

The expression for the third Helmholtz condition for particle $i$ with respect 
to itself in different coordinates $x$ and $y$, analogous to (\ref{eq:HC3_2_1}), reads: 
 \begin{multline}\label{eq:HC3_2_2}
\frac{\partial H_{yi}}{\partial {x_i}} -  \frac{\partial H_{xi}}{\partial {y_i}} +  \frac{\partial }{\partial t}\left(  \frac{\partial H_{xi}}{\partial \dot{y_i}}\right) = \\ = 
\sum_{j} \left[ - T_{ij} \left (\left(L_{i} y_{ij} + d_{i}^{-2}d_{j}^{-1}S_{ij}^{y} \right)\left(\sum_{k}CK_{ik}x_{ik}\right) - \right.\right. \\
\left.\left. \left(L_{i} x_{ij} + d_{i}^{-2}d_{j}^{-1}S_{ij}^{x} \right) \left(\sum_{k}CK_{ik}y_{ik}\right)  \right) - \right.  \\
 \left.-T_{ij}d_{i}^{-1}d_{j}^{-1}\left[vx_{ij}y_{ij} - vy_{ij}x_{ij} \right]\right. \\ 
\left.\left[  \frac{2}{ h\left|r\right|_{ij} \left(1 - 
\frac{\left|r\right|_{ij}}{h}\right)} A - d_{j}^{-1}K_{ij}A  + 
D\left(\left|r\right|_{ij}\right)^{-2}\right]\right] + R^{xy}_{ij}.
 \end{multline}

The expression for the third Helmholtz condition for particle $i$ with respect 
to a different particle $p$ in the same coordinate $x$ 
($y$, $z$), analogous to (\ref{eq:HC3_3_1}), reads: 
\begin{multline}\label{eq:HC3_3_2}
\frac{\partial H_{xp}}{\partial {x_i}} -  \frac{\partial H_{xi}}{\partial {x_p}} +  \frac{\partial }{\partial t}\left(  \frac{\partial H_{xi}}{\partial \dot{x_p}}\right) =  \\ = 
 - C \sum_{j} \left[ T_{ij} \left( K_{ip}x_{ip} \left[  L_{i}x_{ij}  +     d_{i}^{-2}d_{j}^{-1}S^{x}_{ij} \right] -  K_{pj}x_{pj}\left[    L_{j}x_{ij}+ d_{i}^{-1}d_{j}^{-2}S^{x}_{ij} \right]   \right) + \right. \\
\left. + T_{ip} K_{ij}x_{ij} \left( L_{i}x_{ip} + d_{i}^{-2}d_{p}^{-1}S^{x}_{ip} \right) \right]  - \\
- C \sum_{l} \left(  T_{pl} \left[  K_{ip}x_{ip} \left(  L_{p}x_{pl} +  d_{p}^{-2}d_{l}^{-1}S^{x}_{pl}      \right) + K_{il}x_{il} \left(   L_{l}x_{pl} +   d_{p}^{-1}d_{l}^{-2}S^{x}_{pl}   \right)   \right] + \right.\\
 +\left.T_{ip} K_{pl}x_{pl}  \left( L_{p}x_{ip} + d_{i}^{-1}d_{p}^{-2}S^{x}_{ip} 
 \right)\right) + R^{x}_{ip}.
 \end{multline}

The expressions for the third Helmholtz condition for particle  $i$ with respect to a different particle $p$ in different 
coordinates $x$ and $y$, analogous to (\ref{eq:HC3_4_1}) read:
  \begin{multline}\label{eq:HC3_4_2}
\frac{\partial H_{y_p}}{\partial {x_i}} -  \frac{\partial H_{x_i}}{\partial {y_p}} +  \frac{\partial }{\partial t}\left(  \frac{\partial H_{x_i}}{\partial \dot{y_p}}\right) =\\ = 
 -T_{ip} d_{i}^{-1}d_{p}^{-1}\left[ vy_{ip}x_{ip}-vx_{ip}y_{ip}\right]\left[\frac{2}{h\left|r\right|_{ip} \left(1 - \frac{\left|r\right|_{ip}}{h}\right)} A + D\left(\left|r\right|_{ip}\right)^{-2}\right] - \\
 - C\sum_{j} \left[ T_{ij} \left(K_{ip}y_{ip}\left[ L_{i}x_{ij} + d_{i}^{-2}d_{j}^{-1}S_{ij}^{x}\right] - K_{pj}y_{pj}\left[ L_{j}x_{ij} + d_{i}^{-1}d_{j}^{-2}S_{ij}^{x}\right] \right) - \right. \\
 \left. - T_{ip} K_{ij}x_{ij} \left(L_{i} y_{ip} + d_{i}^{-2}d_{p}^{-1}S_{ip}^{y} \right) \right] -\\
 - C\sum_{l} \left[ T_{pl} \left(K_{ip}x_{ip}\left[ L_{p}y_{pl} + d_{p}^{-2}d_{l}^{-1}S_{pl}^{y}\right]  + K_{il}x_{il}\left[ L_{l}y_{pl} + d_{p}^{-1}d_{l}^{-2}S_{pl}^{y}\right]\right) - \right. \\
 \left. - T_{ip} K_{pl}y_{pl}\left(L_{p} x_{ip} + 
d_{i}^{-1}d_{p}^{-2}S_{ip}^{x} \right) \right] +  R^{xy}_{ip},
 \end{multline} 
 \begin{multline}\label{eq:HC3_4_3}
  \frac{\partial H_{xp}}{\partial {y_i}} -  \frac{\partial H_{yi}}{\partial {x_p}} +  \frac{\partial }{\partial t}\left(  \frac{\partial H_{yi}}{\partial \dot{x_p}}\right) = \\ = 
  T_{ip} d_{i}^{-1}d_{p}^{-1}\left[ vy_{ip}x_{ip}-vx_{ip}y_{ip}\right]\left[\frac{2}{h\left|r\right|_{ip}\left(1 - \frac{\left|r\right|_{ip}}{h}\right)} A + D\left(\left|r\right|_{ip}\right)^{-2}\right] + \\
  +  C\sum_{j} \left[T_{ij} \left(K_{pj}x_{pj}\left[ L_{j}y_{ij} + d_{i}^{-1}d_{j}^{-2}S_{ij}^{y}\right] - K_{ip}x_{ip}\left[ L_{i}y_{ij} + d_{i}^{-2}d_{j}^{-1}S_{ij}^{y}\right] \right) \right. \\
  \left. + T_{ip} K_{ij}y_{ij} \left(L_{i} x_{ip} + d_{i}^{-2}d_{p}^{-1}S_{ip}^{x} \right)\right] - \\
-C\sum_{l} \left[ T_{pl} \left( K_{ip}y_{ip}\left[ L_{p}x_{pl} + d_{p}^{-2}d_{l}^{-1}S_{pl}^{x}\right]   \right) - K_{il}y_{il}\left[ L_{l}x_{pl} + d_{p}^{-1}d_{l}^{-2}S_{pl}^{x}\right] \right. - \\
\left. - T_{ip} K_{pl}x_{pl}  \left(L_{p} y_{ip} + 
d_{i}^{-1}d_{p}^{-2}S_{ip}^{y} \right) \right]   +  R^{xy}_{ip},
\end{multline}
\begin{multline}\label{eq:HC3_4_4}
\frac{\partial H_{y_i}}{\partial {x_p}} -  \frac{\partial H_{x_p}}{\partial {y_i}} +  \frac{\partial }{\partial t}\left(  \frac{\partial H_{x_p}}{\partial \dot{y_i}}\right) = \\ = 
- T_{ip} d_{i}^{-1}d_{p}^{-1}\left[ vy_{ip}x_{ip}-vx_{ip}y_{ip}\right]\left[\frac{2}{h\left|r\right|_{ip}\left(1 - \frac{\left|r\right|_{ip}}{h}\right)} A + D\left(\left|r\right|_{ip}\right)^{-2}\right] - \\
- C\sum_{j} \left[T_{ij} \left(K_{pj}x_{pj}\left[ L_{j}y_{ij} + d_{i}^{-1}d_{j}^{-2}S_{ij}^{y}\right] - K_{ip}x_{ip}\left[ L_{i}y_{ij} + d_{i}^{-2}d_{j}^{-1}S_{ij}^{y}\right] \right) \right. \\
\left. + T_{ip} K_{ij}y_{ij} \left(L_{i} x_{ip} + d_{i}^{-2}d_{p}^{-1}S_{ip}^{x} \right)\right] + \\
+C\sum_{l} \left[ T_{pl} \left( K_{ip}y_{ip}\left[ L_{p}x_{pl} + d_{p}^{-2}d_{l}^{-1}S_{pl}^{x}\right]   \right) - K_{il}y_{il}\left[ L_{l}x_{pl} + d_{p}^{-1}d_{l}^{-2}S_{pl}^{x}\right] \right. - \\
\left. - T_{ip} K_{pl}x_{pl}  \left(L_{p} y_{ip} + d_{i}^{-1}d_{p}^{-2}S_{ip}^{y} \right) \right]   
 +  R^{xy}_{ip},
 \end{multline}
where $A = \left( \frac{5 \eta}{3} - \xi \right),$ $D =  \left( \xi + 
\frac{\eta}{3} \right),$ $C =  \frac{315}{4 \pi h^{5}}$  are constants and the following auxiliary variables were 
introduced:
\begin{equation}
L_{i} = 5 B \left( \frac{m_i}{\rho_{0}} \right)^{7} d_{i}^{4}+2 
Bd_{i}^{-3},
\end{equation}
\begin{equation}
S^{x}_{ij}= Avx_{ij} + D x_{ij} \left( r_{ij} \cdot v_{ij} \right) \left( \left| r \right|_{ij} \right)^{-2},
\end{equation}
\begin{multline}
R^{x}_{ij}= \sum_{j}T_{ij}\left[ Q_{ij}^{x} \left( -\frac{2}{h \left| r \right|_{ij} d_i d_j \left(1 - \frac{\left| r \right|_{ij}}{h} \right)}  \left(r_{ij} \cdot v_{ij} \right)   +\right. \right. \\
\left.\left. + d_{i}^{-1}d_{j}^{-2}  \left(\sum_{k}CK_{jk}\left(r_{jk} \cdot v_{jk}\right) \right)+ d_{i}^{-2}d_{j}^{-1}  \left(\sum_{k}CK_{ik}\left(r_{ik} \cdot v_{ik}\right)\right) \right) +  \right. \\
\left. +2 d_{i}^{-1}d_{j}^{-1}  D\left(\left| r \right|_{ij}\right)^{-2} \left[vx_{ij}x_{ij} - \left(r_{ij} \cdot v_{ij} \right) x_{ij}^2\left(\left| r \right|_{ij}\right)^{-2} \right] \right],
 \end{multline}
 \begin{multline}
R^{xy}_{ij}= \sum_{j} T_{ij}\left[ Q^{xy}_{ij} \left(-\frac{2}{h \left| r \right|_{ij} d_i d_j \left(1 - \frac{\left| r \right|_{ij}}{h} \right)}  \left(r_{ij} \cdot v_{ij} \right)     + \right.\right. \\ 
 \left.\left. + d_{i}^{-2}d_{j}^{-1}  \left(\sum_{k}CK_{ik}\left(r_{ik} \cdot v_{ik}\right) \right)+   d_{i}^{-1}d_{j}^{-2}  \left(\sum_{k}CK_{jk}\left(r_{jk} \cdot v_{jk}\right)\right)  \right)  + \right. \\ 
 \left.\left. + d_{i}^{-1}d_{j}^{-1}   
D\left(\left|r\right|_{ij}\right)^{-2}\left[ vx_{ij}y_{ij}+vy_{ij}x_{ij} - 
2\left(\left|r\right|_{ij}\right)^{-2} \left(r_{ij} \cdot v_{ij} \right) x_{ij} 
y_{ij} \right] \right]\right. ,
 \end{multline}
\begin{multline}
R^{x}_{ip}=
T_{ip}\left[ Q^{x}_{ip} \left(\frac{2}{h \left| r \right|_{ip} d_i d_p \left(1 -
\frac{\left| r \right|_{ip}}{h} \right)}  \left(r_{ip} \cdot v_{ip} \right)
- \right. \right.\\ 
\left. - d_{i}^{-1}d_{p}^{-2}  \left(\sum_{j}CK_{pj}\left(r_{pj}
\cdot v_{pj}\right) \right)-   d_{i}^{-2}d_{p}^{-1}
\left(\sum_{j}CK_{ij}\left(r_{ij} \cdot v_{ij}\right)\right)  \right)  +  \\ 
\left. + d_{i}^{-1}d_{p}^{-1}  2 D x_{ip}\left( \left| r \right|_{ip}
\right)^{-2}  \left[ x_{ip} \left( \left| r \right|_{ip} \right)^{-2}
\left(r_{ip} \cdot v_{ip} \right) - vx_{ip} \right]   \right],
 \end{multline}
 \begin{multline}
R^{xy}_{ip}=  
T_{ip}\left[ Q^{xy}_{ip} \left(-\frac{2}{h \left| r \right|_{ip} d_i d_p \left(1
- \frac{\left| r \right|_{ip}}{h} \right)}  \left(r_{ip} \cdot v_{ip} \right)
+ \right. \right.\\ 
\left. + d_{i}^{-1}d_{p}^{-2}  \left(\sum_{j}CK_{pj}\left(r_{pj}
\cdot v_{pj}\right) \right)+   d_{i}^{-2}d_{p}^{-1}
\left(\sum_{j}CK_{ij}\left(r_{ij} \cdot v_{ij}\right)\right)  \right)  + \\ 
\left. + d_{i}^{-1}d_{p}^{-1}  D\left(\left| r
\right|_{ip}\right)^{-2}\left[vx_{ip}y_{ip} +x_{ip}vy_{ip} -2\left(r_{ip} \cdot
v_{ip} \right) x_{ip}y_{ip}\left(\left| r \right|_{ip}\right)^{-2} \right]
\right].
\label{eq:last}
 \end{multline}

\section{Macroscopic limit for the second and the third Helmholtz conditions}
\label{app:limits}

The same macroscopic-limit substitutions as in Sec.~III are used:
\begin{equation}
h=G N^{-1/3},\qquad d_i=E N,\qquad m_i=\frac{M_0}{N},
\end{equation}
\begin{equation}
|\mathbf r_{ij}|=\left(\frac{j}{K_0}\right)^{1/3}h,
\qquad
K_{ij}=\left(1-\left(\frac{j}{K_0}\right)^{1/3}\right)^2,
\end{equation}
\begin{equation}
x_{ij}=\pm \left(\frac{V_0}{N}\right)^{1/3}j_x,
\qquad
y_{ij}=\pm \left(\frac{V_0}{N}\right)^{1/3}j_y,
\end{equation}
where $j_x$ and $j_y$ are independent $N$. The number of neighbours $K_0$ is fixed and also independent of $N$. Constants independent of $N$ are omitted in the final expressions for simplification purposes.

With the above scalings, the physical mass density is
\begin{equation}
\rho_i=m_i d_i
=
\frac{M_0}{N} E N
=
M_0 E,
\end{equation}
and therefore
\begin{equation}
\rho_i\sim N^0.
\end{equation}
The pressure is evaluated using the physical density, see Sec.~4 below.

\subsection*{1. Second Helmholtz condition: Eq.~(12)}

For particle $i$ with respect to itself in two different coordinates, e.g.\ $x$ and $y$,
\begin{equation}
\frac{\partial H_{x_i}}{\partial \dot y_i}
+
\frac{\partial H_{y_i}}{\partial \dot x_i}
=
\sum_{j=1}^{K_0}
T_{ij}d_i^{-1}d_j^{-1}Q^{xy}_{ij},
\end{equation}
where
\begin{equation}
Q^{xy}_{ij}=D x_{ij}y_{ij}|\mathbf r_{ij}|^{-2}.
\end{equation}

Substituting the $N$-dependences gives
\begin{align}
&\frac{\partial H_{x_i}}{\partial \dot y_i}
+
\frac{\partial H_{y_i}}{\partial \dot x_i}
\notag\\
&=
\frac{315}{4\pi}
\sum_{j=1}^{K_0}
(GN^{-1/3})^{-5}
\left(1-\left(\frac{j}{K_0}\right)^{1/3}\right)^2
M_0^2(NE)^{-2}(N)^{-2}
\notag\\
&\qquad\times
D\left[
\pm V_0^{2/3}N^{-2/3}j_xj_y
\left(\frac{j}{K_0}\right)^{-2/3}
(GN^{-1/3})^{-2}
\right].
\end{align}

The $N$-dependence within the summation simplifies to
\begin{equation}
N^{5/3}N^{-4}N^{-2/3}N^{2/3}
=
N^{-7/3}.
\end{equation}
Thus
\begin{equation}
\frac{\partial H_{x_i}}{\partial \dot y_i}
+
\frac{\partial H_{y_i}}{\partial \dot x_i}
\sim
\sum_{j=1}^{K_0}
N^{-7/3}
\end{equation}
and the residue tends to $0$ as $N\to\infty$. This result also holds for the coordinate pairs $(x,z)$ and $(y,z)$.

\subsection*{2. Second Helmholtz condition: Eq.~(13)}

For particle $i$ with respect to a different particle $p$ in the same coordinate, e.g.\ $x$,
\begin{equation}
\frac{\partial H_{x_i}}{\partial \dot x_p}
+
\frac{\partial H_{x_p}}{\partial \dot x_i}
=
T_{ip}d_i^{-1}d_p^{-1}Q^x_{ip},
\end{equation}
where
\begin{equation}
Q^x_{ip}=A+D x_{ip}|\mathbf r_{ip}|^{-2}.
\end{equation}

Substituting the $N$-dependences gives
\begin{align}
&\frac{\partial H_{x_i}}{\partial \dot x_p}
+
\frac{\partial H_{x_p}}{\partial \dot x_i}
\notag\\
&=
\frac{315}{4\pi}
(GN^{-1/3})^{-5}
\left(1-\left(\frac{p}{K_0}\right)^{1/3}\right)^2
M_0^2(NE)^{-2}(N)^{-2}
\notag\\
&\qquad\times
\left[
A
\pm
D V_0^{1/3}N^{-1/3}p
\left(\frac{p}{K_0}\right)^{-2/3}
(GN^{-1/3})^{-2}
\right].
\end{align}
The two contributions have the following $N$-dependences:
\begin{equation}
N^{5/3}N^{-4}A
\sim
N^{-7/3},
\end{equation}
and
\begin{equation}
N^{5/3}N^{-4}D N^{-1/3}N^{2/3}
\sim
N^{-2}.
\end{equation}
Therefore
\begin{equation}
\frac{\partial H_{x_i}}{\partial \dot x_p}
+
\frac{\partial H_{x_p}}{\partial \dot x_i}
\sim
N^{-7/3}\pm N^{-2}
\end{equation}
and the residue tends to $0$ as $N\to\infty$. This result also holds for the $y$ and $z$ coordinates.

\subsection*{3. Second Helmholtz condition: Eq.~(14)}

For particle $i$ with respect to a different particle $p$ in different coordinates, e.g.\ $x$ and $y$,
\begin{equation}
\frac{\partial H_{x_i}}{\partial \dot y_p}
+
\frac{\partial H_{y_p}}{\partial \dot x_i}
=
T_{ip}d_i^{-1}d_p^{-1}Q^{xy}_{ip},
\end{equation}
where
\begin{equation}
Q^{xy}_{ip}=D x_{ip}y_{ip}|\mathbf r_{ip}|^{-2}.
\end{equation}

Substituting the $N$-dependences gives
\begin{align}
&\frac{\partial H_{x_i}}{\partial \dot y_p}
+
\frac{\partial H_{y_p}}{\partial \dot x_i}
\notag\\
&=
\frac{315}{4\pi}
(GN^{-1/3})^{-5}
\left(1-\left(\frac{p}{K_0}\right)^{1/3}\right)^2
M_0^2(NE)^{-2}(N)^{-2}
\notag\\
&\qquad\times
D\left[
\pm V_0^{2/3}N^{-2/3}p_xp_y
\left(\frac{p}{K_0}\right)^{-2/3}
(GN^{-1/3})^{-2}
\right].
\end{align}

This simplifies to
\begin{equation}
N^{5/3}N^{-4}N^{-2/3}N^{2/3}
=
N^{-7/3}.
\end{equation}
Thus
\begin{equation}
\frac{\partial H_{x_i}}{\partial \dot y_p}
+
\frac{\partial H_{y_p}}{\partial \dot x_i}
\sim
N^{-7/3}
\end{equation}
and the residue tends to $0$ as $N\to\infty$. This result also holds for the other coordinate pairs.

\subsection*{4. Auxiliary $N$-dependences for the third Helmholtz condition}

For the third Helmholtz condition we assume, as in Sec.~III, that the velocity differences scale with $N$ in the same way as the coordinate differences:
\begin{equation}
v_{ij}\sim N^{-1/3}.
\end{equation}
Hence
\begin{equation}
\mathbf r_{ij}\cdot\mathbf v_{ij}\sim N^{-2/3}.
\end{equation}

The frequently occurring terms have the following dependences:
\begin{equation}
C=\frac{315}{4\pi h^5}\sim N^{5/3},
\end{equation}
\begin{equation}
T_{ij}=C K_{ij}m_i m_j\sim N^{5/3}N^{-2}=N^{-1/3},
\end{equation}
\begin{equation}
x_{ij},y_{ij},|\mathbf r_{ij}|\sim N^{-1/3},
\qquad
|\mathbf r_{ij}|^{-2}\sim N^{2/3},
\end{equation}
\begin{equation}
x_{ij}|\mathbf r_{ij}|^{-2}\sim N^{1/3},
\qquad
x_{ij}y_{ij}|\mathbf r_{ij}|^{-2}\sim N^0,
\end{equation}
\begin{equation}
\sum_{k=1}^{K_0} C K_{ik}x_{ik}
\sim
N^{5/3}N^{-1/3}
=
N^{4/3},
\end{equation}
\begin{equation}
\sum_{k=1}^{K_0} C K_{ik}(\mathbf r_{ik}\cdot\mathbf v_{ik})
\sim
N^{5/3}N^{-2/3}
=
N.
\end{equation}

The auxiliary variables $S^x_{ij}$, $S^y_{ij}$, $S^z_{ij}$ have the form
\begin{equation}
S^x_{ij}
=
A v^x_{ij}
+
D x_{ij}(\mathbf r_{ij}\cdot\mathbf v_{ij})|\mathbf r_{ij}|^{-2}.
\end{equation}
The first term is $\sim N^{-1/3}$. The second term is
\begin{equation}
x_{ij}(\mathbf r_{ij}\cdot\mathbf v_{ij})|\mathbf r_{ij}|^{-2}
\sim
N^{-1/3}N^{-2/3}N^{2/3}
=
N^{-1/3}.
\end{equation}
Therefore
\begin{equation}
S^x_{ij},S^y_{ij},S^z_{ij}\sim N^{-1/3}.
\end{equation}

For terms containing $L_i$, we use the equation of state in which the Tait pressure depends on the physical mass density $\rho_i=m_i d_i$:
\begin{equation}
P_i
=
B\left[
\left(\frac{\rho_i}{\rho_0}\right)^7-1
\right]
=
B\left[
\left(\frac{m_i d_i}{\rho_0}\right)^7-1
\right].
\end{equation}

The pressure-dependent part of Eq.~(9) therefore contains
\begin{equation}
B\left(\frac{m_i}{\rho_0}\right)^7 d_i^5
\end{equation}
and the term $B d_i^{-2}$ from the expansion of parentheses in the Tait equation. With
\begin{equation}
m_i=\frac{M_0}{N},
\qquad
d_i=E N,
\end{equation}
the first term becomes
\begin{equation}
B\left(\frac{m_i}{\rho_0}\right)^7 d_i^5
=
B\left(\frac{M_0}{N\rho_0}\right)^7 (E N)^5
=
B E^5\left(\frac{M_0}{\rho_0}\right)^7 N^{-2},
\end{equation}
and the second term becomes
\begin{equation}
-B d_i^{-2}=-B E^{-2}N^{-2}.
\end{equation}
Thus the pressure combination entering Eq.~(9) scales as
\begin{equation}
N^{-2}.
\end{equation}

Accordingly,
\begin{equation}
L_i
=
5B\left(\frac{m_i}{\rho_0}\right)^7 d_i^4
+
2B d_i^{-3}.
\end{equation}
Substituting $m_i=M_0/N$ and $d_i=E N$ gives
\begin{equation}
L_i
=
\left[
5B E^4\left(\frac{M_0}{\rho_0}\right)^7
+
2B E^{-3}
\right]N^{-3}.
\end{equation}
Therefore
\begin{equation}
L_i\sim N^{-3}.
\end{equation}

Thus
\begin{equation}
L_i x_{ij}\sim N^{-3}N^{-1/3}=N^{-10/3},
\end{equation}
and
\begin{equation}
d_i^{-2}d_j^{-1}S^x_{ij}
\sim
N^{-3}N^{-1/3}
=
N^{-10/3}.
\end{equation}
Therefore the combination
\begin{equation}
L_i x_{ij}+d_i^{-2}d_j^{-1}S^x_{ij}
\end{equation}
has the dependence
\begin{equation}
N^{-10/3}.
\end{equation}
This also holds for the $y$ and $z$ components.

\subsection*{5. Third Helmholtz condition: Eq.~(19)}

We now evaluate each term within the residue $R^x_{ij}$ from Eq.~(A8) for a particle $i$ with respect to itself in the same coordinate.

First,
\begin{equation}
Q^x_{ij}=A+D x_{ij}|\mathbf r_{ij}|^{-2}
\sim
A+D N^{1/3}.
\end{equation}

The three terms which are multiplied by $Q^x_{ij}$ in Eq.~(A8) have the following dependences:
\begin{equation}
\frac{1}{h|\mathbf r_{ij}|}
d_i^{-1}d_j^{-1}
(\mathbf r_{ij}\cdot\mathbf v_{ij})
\sim
N^{2/3}N^{-2}N^{-2/3}
=
N^{-2},
\end{equation}
\begin{equation}
d_i^{-1}d_j^{-2}
\sum_{k=1}^{K_0} C K_{jk}(\mathbf r_{jk}\cdot\mathbf v_{jk})
\sim
N^{-3}N
=
N^{-2},
\end{equation}
\begin{equation}
d_i^{-2}d_j^{-1}
\sum_{k=1}^{K_0} C K_{ik}(\mathbf r_{ik}\cdot\mathbf v_{ik})
\sim
N^{-3}N
=
N^{-2}.
\end{equation}

Therefore
\begin{equation}
T_{ij}Q^x_{ij}N^{-2}
\sim
N^{-1/3}(A+D N^{1/3})N^{-2}
\sim
N^{-7/3}\pm N^{-2}.
\end{equation}

The remaining term in Eq.~(A8) is
\begin{equation}
2d_i^{-1}d_j^{-1}
D|\mathbf r_{ij}|^{-2}
\left[
v^x_{ij}x_{ij}
-
(\mathbf r_{ij}\cdot\mathbf v_{ij})
x_{ij}^2|\mathbf r_{ij}|^{-2}
\right].
\end{equation}
Here
\begin{equation}
v^x_{ij}x_{ij}\sim N^{-2/3},
\end{equation}
and
\begin{equation}
(\mathbf r_{ij}\cdot\mathbf v_{ij})
x_{ij}^2|\mathbf r_{ij}|^{-2}
\sim
N^{-2/3}N^{-2/3}N^{2/3}
=
N^{-2/3}.
\end{equation}
Thus
\begin{equation}
|\mathbf r_{ij}|^{-2}
\left[
v^x_{ij}x_{ij}
-
(\mathbf r_{ij}\cdot\mathbf v_{ij})
x_{ij}^2|\mathbf r_{ij}|^{-2}
\right]
\sim
N^{2/3}N^{-2/3}
=
N^0.
\end{equation}
Multiplying by $T_{ij}d_i^{-1}d_j^{-1}$ gives
\begin{equation}
T_{ij}d_i^{-1}d_j^{-1}
\sim
N^{-1/3}N^{-2}
=
N^{-7/3}.
\end{equation}

Hence the residue in Eq.~(19) has the dependence
\begin{equation}
R^x_{ij}
\sim
\sum_{j=1}^{K_0}
\left(
N^{-7/3}\pm N^{-2}
\right)
\end{equation}
and therefore tends to $0$ as $N\to\infty$. This result also holds for the $y$ and $z$ coordinates.

\subsection*{6. Third Helmholtz condition: Eq.~(20) and Eq.~(A1)}

For particle $i$ with respect to itself in different coordinates, e.g.\ $x$ and $y$, Eq.~(20) contains three distinct terms.

The first is
\begin{equation}
T_{ij}
\left(
L_i y_{ij}
+
d_i^{-2}d_j^{-1}S^y_{ij}
\right)
\sum_{k=1}^{K_0} C K_{ik}x_{ik}.
\end{equation}
Using
\begin{equation}
L_i y_{ij}
+
d_i^{-2}d_j^{-1}S^y_{ij}
\sim
N^{-10/3},
\end{equation}
and
\begin{equation}
\sum_{k=1}^{K_0} C K_{ik}x_{ik}
\sim
N^{4/3},
\end{equation}
we obtain
\begin{equation}
T_{ij}N^{-10/3}N^{4/3}
\sim
N^{-1/3}N^{-2}
=
N^{-7/3}.
\end{equation}

The second is
\begin{equation}
T_{ij}d_i^{-1}d_j^{-1}
\left[
v^x_{ij}y_{ij}-v^y_{ij}x_{ij}
\right]
\left[
\frac{2}{h|\mathbf r_{ij}|}
\left(1-\frac{|\mathbf r_{ij}|}{h}\right)A
-
d_j^{-1}K_{ij}
\left(
A+D|\mathbf r_{ij}|^{-2}
\right)
\right].
\end{equation}
The velocity factor scales as
\begin{equation}
v^x_{ij}y_{ij}-v^y_{ij}x_{ij}
\sim
N^{-2/3}.
\end{equation}
The first part of the second bracket gives
\begin{equation}
\frac{2}{h|\mathbf r_{ij}|}A
\sim
N^{2/3}.
\end{equation}
The second part gives
\begin{equation}
d_j^{-1}K_{ij}
\left(
A+D|\mathbf r_{ij}|^{-2}
\right)
\sim
N^{-1}(A+D N^{2/3})
\sim
N^{-1/3}.
\end{equation}
Therefore this term has the dependence
\begin{equation}
T_{ij}d_i^{-1}d_j^{-1}
N^{-2/3}
\left(
N^{2/3}+N^{-1/3}
\right)
\sim
N^{-7/3}+N^{-10/3}.
\end{equation}

The remaining contribution is $R^{xy}_{ij}$, Eq.~(A9). In this expression,
\begin{equation}
Q^{xy}_{ij}=D x_{ij}y_{ij}|\mathbf r_{ij}|^{-2}\sim N^0.
\end{equation}
The terms multiplied by $Q^{xy}_{ij}$ contain the same factors as in Eq.~(A8) and are $\sim N^{-2}$. Hence
\begin{equation}
T_{ij}Q^{xy}_{ij}N^{-2}
\sim
N^{-1/3}N^{-2}
=
N^{-7/3}.
\end{equation}

The last term in Eq.~(A9) contains
\begin{equation}
|\mathbf r_{ij}|^{-2}
\left[
v^x_{ij}y_{ij}+v^y_{ij}x_{ij}
-
2(\mathbf r_{ij}\cdot\mathbf v_{ij})
x_{ij}y_{ij}|\mathbf r_{ij}|^{-2}
\right].
\end{equation}
The expression within the square brackets simplifies to $\sim N^{-2/3}$, and $|\mathbf r_{ij}|^{-2}\sim N^{2/3}$, so the product is $\sim N^0$. Multiplying by $T_{ij}d_i^{-1}d_j^{-1}$ gives
\begin{equation}
N^{-1/3}N^{-2}=N^{-7/3}.
\end{equation}
Thus
\begin{equation}
R^{xy}_{ij}
\sim
\sum_{j=1}^{K_0}
N^{-7/3}.
\end{equation}

Collecting the terms, we obtain
\begin{equation}
\text{Eq.~(20)}
\sim
\sum_{j=1}^{K_0}
\left(
N^{-7/3}+N^{-10/3}
\right)
\end{equation}
and therefore the residue tends to $0$ as $N\to\infty$.

Equation (A1) is identical Eq.~(20) apart from sign differences, therefore
\begin{equation}
\text{Eq.~(A1)}
\sim
\sum_{j=1}^{K_0}
\left(
N^{-7/3}+N^{-10/3}
\right)
\end{equation}
and also tends to $0$. This result also holds for the other coordinate pairs.

\subsection*{7. Third Helmholtz condition: Eq.~(21) and Eq.~(A2)}

For particle $i$ with respect to a different particle $p$ in the same coordinate, e.g.\ $x$, Eq.~(21):

A generic term in the first sum has the form
\begin{equation}
C T_{ij}K_{ip}x_{ip}
\left(
L_i x_{ij}
+
d_i^{-2}d_j^{-1}S^x_{ij}
\right).
\end{equation}
Using
\begin{equation}
C\sim N^{5/3},
\qquad
T_{ij}\sim N^{-1/3},
\qquad
x_{ip}\sim N^{-1/3},
\end{equation}
and
\begin{equation}
L_i x_{ij}
+
d_i^{-2}d_j^{-1}S^x_{ij}
\sim
N^{-10/3},
\end{equation}
we obtain
\begin{equation}
C T_{ij}x_{ip}N^{-10/3}
\sim
N^{5/3}N^{-1/3}N^{-1/3}N^{-10/3}
=
N^{-7/3}.
\end{equation}

The second term inside the first sum,
\begin{equation}
C T_{ip}K_{ij}x_{ij}
\left(
L_i x_{ip}
+
d_i^{-2}d_p^{-1}S^x_{ip}
\right),
\end{equation}
has the same dependence,
\begin{equation}
N^{-7/3}.
\end{equation}

The second sum over $l$ contains analogous terms and therefore also contributes terms with dependence
\begin{equation}
N^{-7/3}.
\end{equation}

The residue $R^x_{ip}$, Eq.~(A10), is evaluated in the same way. The terms multiplied by $Q^x_{ip}$ are $\sim N^{-2}$, while
\begin{equation}
Q^x_{ip}\sim A+D N^{1/3}.
\end{equation}
Thus
\begin{equation}
T_{ip}Q^x_{ip}N^{-2}
\sim
N^{-7/3}\pm N^{-2}.
\end{equation}

The final term in Eq.~(A10) gives
\begin{equation}
T_{ip}d_i^{-1}d_p^{-1}
\sim
N^{-1/3}N^{-2}
=
N^{-7/3}.
\end{equation}
Therefore
\begin{equation}
R^x_{ip}
\sim
N^{-7/3}\pm N^{-2}.
\end{equation}

Collecting terms,
\begin{equation}
\text{Eq.~(21)}
\sim
\sum_{j=1}^{K_0}N^{-7/3}
+
\sum_{l=1}^{K_0}N^{-7/3}
+
N^{-7/3}\pm N^{-2}
\end{equation}
and hence tends to $0$ as $N\to\infty$.

Equation (A2) is identical Eq.~(21) apart from sign differences for the summation terms, therefore
\begin{equation}
\text{Eq.~(A2)}
\sim
\sum_{j=1}^{K_0}N^{-7/3}
+
\sum_{l=1}^{K_0}N^{-7/3}
+
N^{-7/3}\pm N^{-2}
\end{equation}
and also tends to $0$. This result also holds for the $y$ and $z$ coordinates.

\subsection*{8. Third Helmholtz condition: Eq.~(22) and Eqs.~(A3)--(A5)}

For particle $i$ with respect to a different particle $p$ in different coordinates, e.g.\ $x$ and $y$, Eq.~(22):

The first term we examine is
\begin{equation}
T_{ip}d_p^{-1}
\left[
v^y_{ip}x_{ip}-v^x_{ip}y_{ip}
\right]
\left[
\frac{2}{h|\mathbf r_{ip}|}
\left(1-\frac{|\mathbf r_{ip}|}{h}\right)A
+
D|\mathbf r_{ip}|^{-2}
\right].
\end{equation}
We can break down the individual factors:
\begin{equation}
T_{ip}\sim N^{-1/3},
\qquad
d_p^{-1}\sim N^{-1},
\end{equation}
\begin{equation}
v^y_{ip}x_{ip}-v^x_{ip}y_{ip}
\sim
N^{-2/3},
\end{equation}
and
\begin{equation}
\frac{2}{h|\mathbf r_{ip}|}A
+
D|\mathbf r_{ip}|^{-2}
\sim
N^{2/3}.
\end{equation}
Therefore the term simplifies down to
\begin{equation}
N^{-1/3}N^{-1}N^{-2/3}N^{2/3}
=
N^{-4/3}.
\end{equation}

The terms within the two summations have the same structure as in Eq.~(21), namely
\begin{equation}
C T K x
\left(
Lx+d^{-2}d^{-1}S
\right),
\end{equation}
and each such term has the dependence
\begin{equation}
N^{-7/3}.
\end{equation}

The residue $R^{xy}_{ip}$, Eq.~(A11), is evaluated in the same way. Since
\begin{equation}
Q^{xy}_{ip}\sim N^0,
\end{equation}
and the terms multiplied by $Q^{xy}_{ip}$ are $\sim N^{-2}$, we obtain
\begin{equation}
T_{ip}Q^{xy}_{ip}N^{-2}
\sim
N^{-1/3}N^{-2}
=
N^{-7/3}.
\end{equation}

The final term in Eq.~(A11) gives
\begin{equation}
T_{ip}d_i^{-1}d_p^{-1}
\sim
N^{-7/3}.
\end{equation}
Therefore
\begin{equation}
R^{xy}_{ip}
\sim
N^{-7/3}.
\end{equation}

Thus Eq.~(22) has the dependence
\begin{equation}
\text{Eq.~(22)}
\sim
N^{-4/3}
+
\sum_{j=1}^{K_0}N^{-7/3}
+
\sum_{l=1}^{K_0}N^{-7/3}
+
N^{-7/3}
\end{equation}
and tends to $0$ as $N\to\infty$.

Equation (A3) has the same structure as Eq.~(22), with reversed signs. Therefore,
\begin{equation}
\text{Eq.~(A3)}
\sim
N^{-4/3}
+
\sum_{j=1}^{K_0}N^{-7/3}
+
\sum_{l=1}^{K_0}N^{-7/3}
+
N^{-7/3}
\end{equation}
and tends to $0$.

Equation (A4) also has the same dependence:
\begin{equation}
\text{Eq.~(A4)}
\sim
N^{-4/3}
+
\sum_{j=1}^{K_0}N^{-7/3}
+
\sum_{l=1}^{K_0}N^{-7/3}
+
N^{-7/3}
\end{equation}
and tends to $0$.

Equation (A5) differs in that the first term contains two inverse density factors:
\begin{equation}
T_{ip}d_i^{-1}d_p^{-1}
\left[
v^y_{ip}x_{ip}-v^x_{ip}y_{ip}
\right]
\left[
\frac{2}{h|\mathbf r_{ip}|}
\left(1-\frac{|\mathbf r_{ip}|}{h}\right)A
+
D|\mathbf r_{ip}|^{-2}
\right].
\end{equation}
The corresponding dependence is
\begin{equation}
N^{-1/3}N^{-2}N^{-2/3}N^{2/3}
=
N^{-7/3}.
\end{equation}

The summation terms and $R^{xy}_{ip}$ again contribute $N^{-7/3}$, and thus
\begin{equation}
\text{Eq.~(A5)}
\sim
\sum_{j=1}^{K_0}N^{-7/3}
+
\sum_{l=1}^{K_0}N^{-7/3}
+
N^{-7/3}
\end{equation}
and also tends to $0$.

The same dependences hold for the other coordinate combinations.

\end{widetext}

\end{appendix}

\end{document}